\documentclass[11pt]{article}
\usepackage{epsfig}

\newcommand {\ms} {M$_S$}

\begin{document}
\newcommand{\Dzero}{D\O\ } \title{Probing Extra Spacetime Dimensions
at the Tevatron} \author{{\large Laurent Duflot$^1$ and Marumi
Kado$^2$} \vspace{.2cm}
\\ {\it On Behalf of the CDF and \Dzero Collaborations} \\
{\it Talk presented at the SUSY '02 conference, DESY Hamburg} \\
\\ {\small {\it $^1$Laboratoire de l'Acc\'el\'erateur Lin\'eaire,}}
\\{\small {\it Universit\'e de Paris-Sud, 91898 Orsay Cedex, France}} \vspace{.3cm}
\\ {\small {\it $^2$Lawrence Berkeley National Laboratory,}}
\\ {\small {\it Berkeley, California 94720, USA}}}
\date{}
\maketitle
\vspace{-.8cm}
\parbox{15cm}{\abstract{Theories with extra spacetime dimensions
aiming at resolving the hierarchy problem have recently been
developed. These scenarios have provided exciting new grounds for
experimental probes. A review of the searches conducted at the
Tevatron in this framework during its first running phase and the
prospects for its second running phase are reviewed.}}


\vspace{.5cm}
\noindent{\bf {\large Introduction}}
\vspace{.5cm}

Among the most popular scenarios of theories with extra spacetime
dimensions, those from Arkani-Hamed, Dimopoulos and Dvali (ADD)~\cite{add} and
Randall-Sundrum (RS)~\cite{rs1} have provided exciting benchmark
signatures to be searched for at colliders and in particular at the
Tevatron.

The Tevatron program started its data taking period in 1992 at a
center-of-mass energy of 1.8\,TeV. This first running (Run\,1) phase
extended until 1996, reaching a peak luminosity of 2$\times 10
^{30}$~cm$^{-2}$s$^{-1}$ and an integrated luminosity of about
100\,pb$^{-1}$ for each experiment, CDF and \Dzero. In spring 2001,
after a complete reappraisal of the machine and the detectors, the
Tevatron started taking data again but at a higher center-of-mass
energy of 1.96\,TeV. The booster, accumulator ring and the cooling
were upgraded but the most dramatic change was the construction of a
main injector along with an anti-proton recycler which where built in
an tunnel adjacent to the Tevatron ring.  Run\,2 will be divided into
two periods, Run\,2a and Run\,2b, in between which further upgrades
are foreseen. The design peak luminosities for these two running
periods are respectively $\sim$10$^{32}$ and
$\sim$5$\times$10$^{32}$~cm$^{-2}$s$^{-1}$, aiming at integrated
luminosities per experiment of 2\,fb$^{-1}$ and 15\,fb$^{-1}$
respectively. The detectors underwent extensive changes. CDF has a new
inner silicon tracker and plug calorimeter, upgraded muon detectors
and a new trigger and data acquisition system. \Dzero\ installed a new
superconducting 2\,Teslas solenoid in which a complete new tracker was
built. This tracker is a combination of a silicon tracker and a
scintillating fibers detector. \Dzero\ as well has a completely new
trigger and data acquisition system.

\newpage

\vspace{.5cm}
\noindent{\bf {\large The Large Extra Dimension Scenario (ADD)}}
\vspace{.5cm}

\begin{figure}[t] 
\begin{center}
\begin{tabular}{ccc}
\epsfig{file=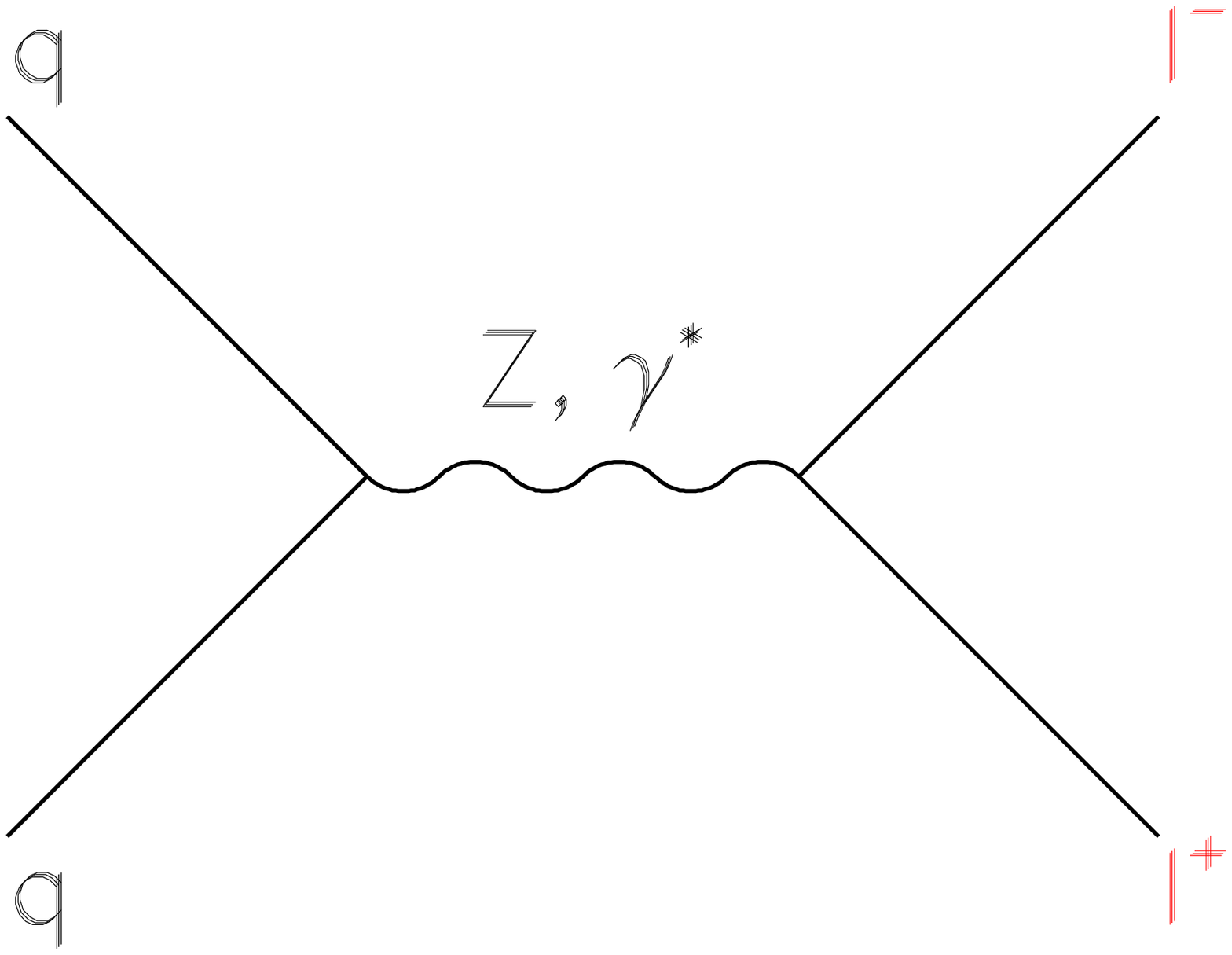, width=3.3cm}&
\epsfig{file=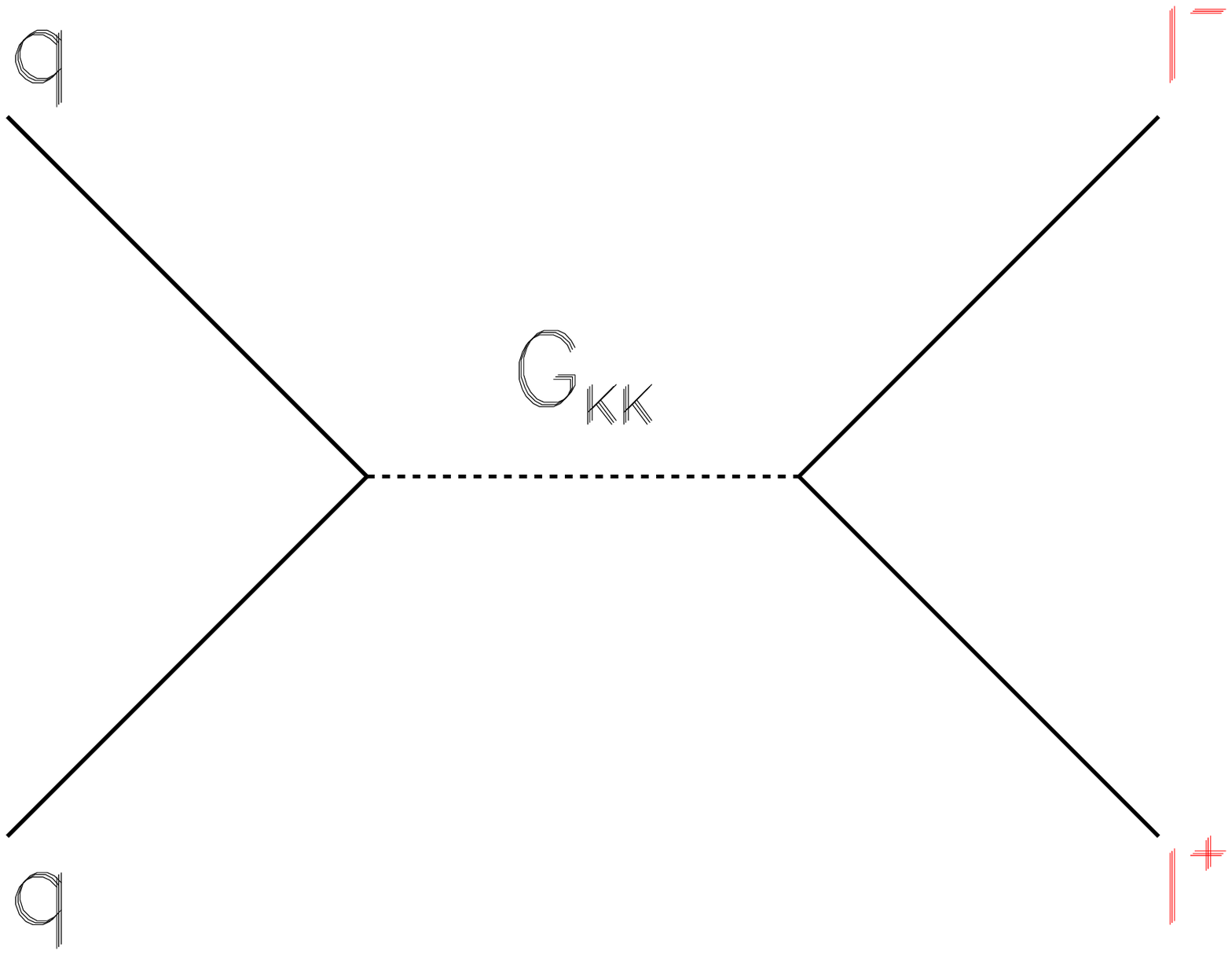, width=3.3cm}&
\epsfig{file=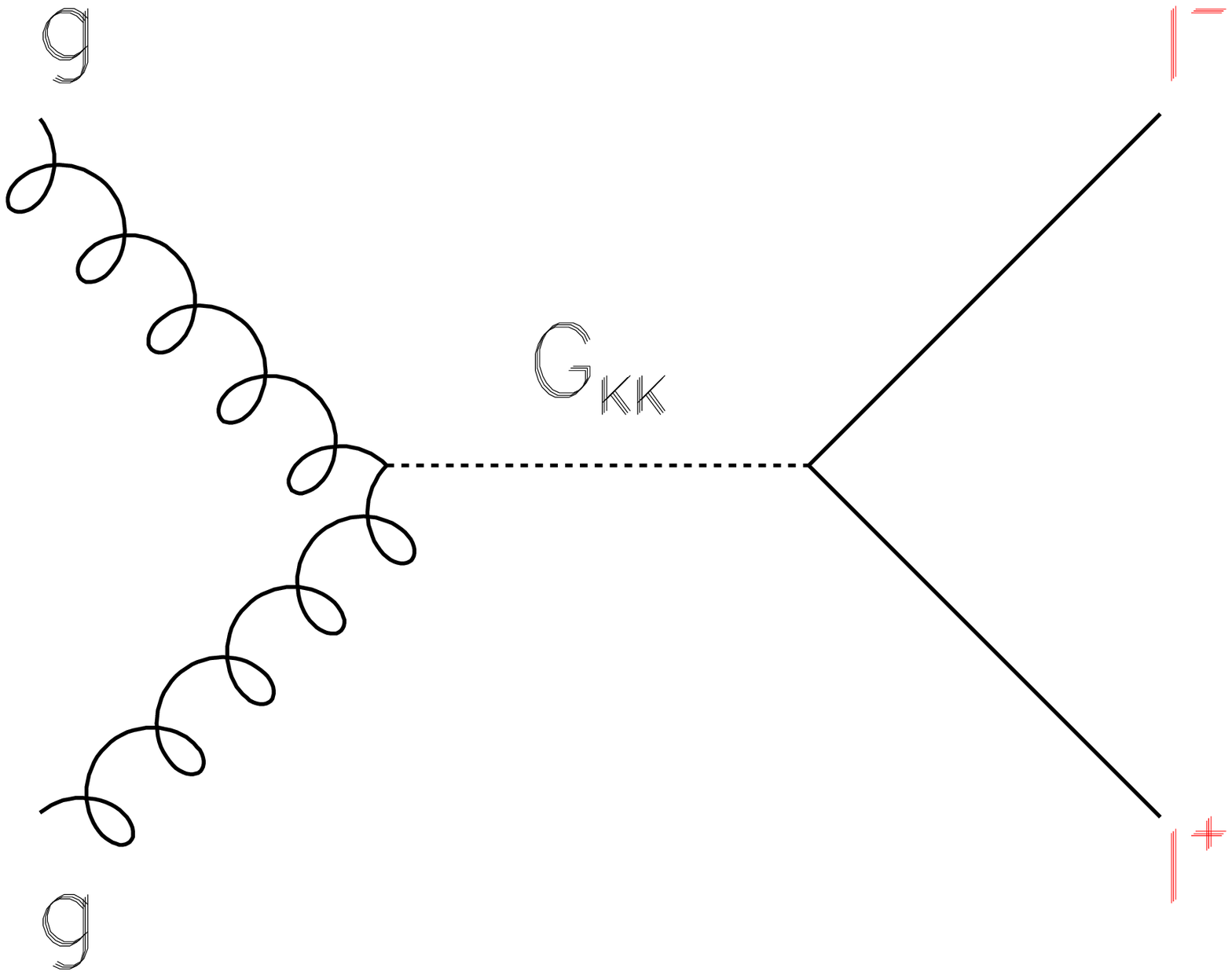, width=3.3cm}
\end{tabular}
  \parbox{15cm}{\caption{{\small Diagrams of the standard model (left) and KK graviton
  contributions (middle and right) to di-lepton production.}}
  \label{dilepton}}
\end{center}
\end{figure}

In the ADD scenario, the observed hierarchy between the TeV scale and
the Planck scale arises from the adjunction of $n$ additional
compactified dimensions in which only gravity can propagate. The
Planck scale in the 4-d theory $M_{Pl}$ is related to the gravity
scale  $M_S$ in the 4+$n$ dimensional spacetime via Gauss' theorem
through:
\begin{equation}
M_{Pl}^2=M_S^{n+2}\times (2\pi R_c)^n
\end{equation}	
Here flat compactified dimensions of toroidal form and equal size are
considered. The hierarchy problem is then solved when extra dimensions
are large (from one millimeter to $\sim$10\,fm for $n$ ranging between
2 and 6, the $n=1$ case is excluded as it would affect Newton's law at
large distances). In the 4-d world, excitations of the states
propagating in the bulk (called Kaluza-Klein modes) appear as a
consequence of the compactification of the extra dimensions. In the
simplest models, these states are equidistant and their separation is
inversely proportional to the size of the extra dimensions. Due to the
large compactification radius, the number of accessible Kaluza-Klein (KK)
modes at a collider is large enough to compensate for the Planck scale
suppression of each individual KK state coupling and the spectrum of
KK modes would appear to be almost continuous given the detector
resolutions. The presence of extra dimensions in these scenarios thus
results in two generic types of signatures: the alteration of
production cross sections and asymmetries of processes such
$q\overline{q},gg\rightarrow \ell^+\ell^-, \gamma\gamma$ ({\it n.b.} in
this framework, the $gg\rightarrow \ell^+\ell^-$ is a pure extra
dimension process) and the direct KK graviton emission in association
with a vector boson.

\vspace{.5cm}
\noindent{\it Virtual Kaluza-Klein Graviton Exchange}
\vspace{.5cm}

\begin{figure}[t] 
\begin{center}
\begin{tabular}{cccc}
\epsfig{file=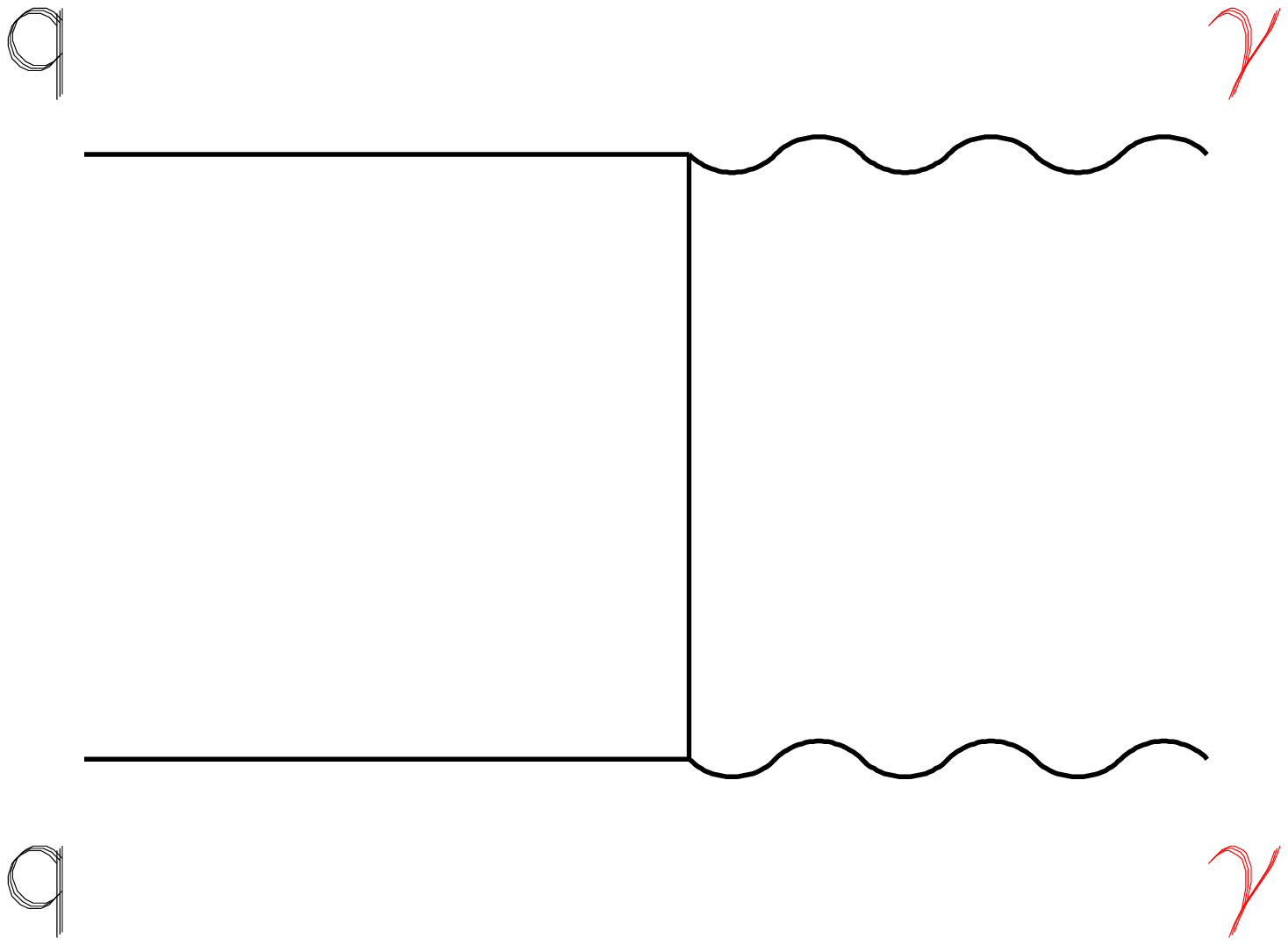, width=3.3cm}&
\epsfig{file=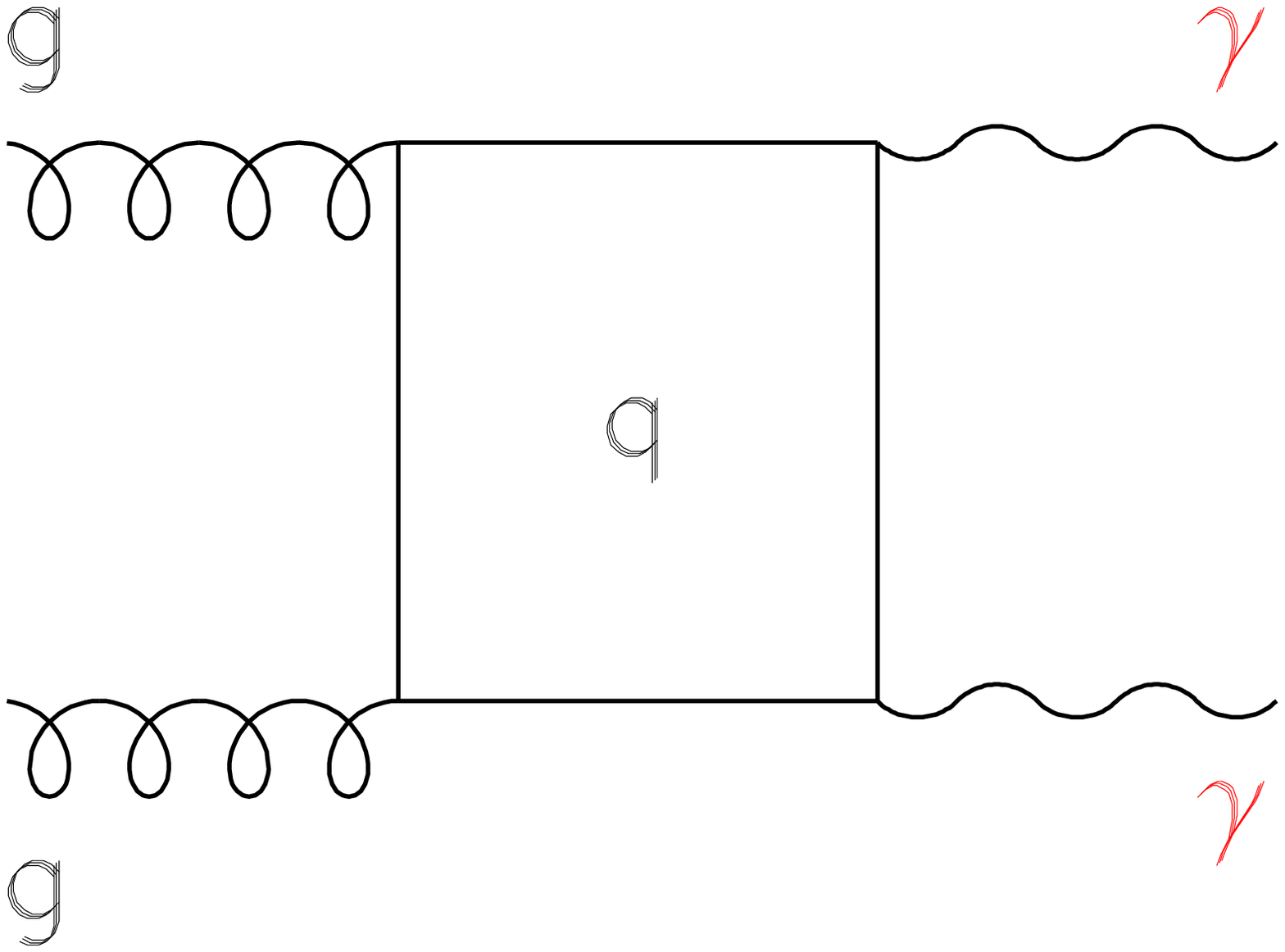, width=3.3cm}&
\epsfig{file=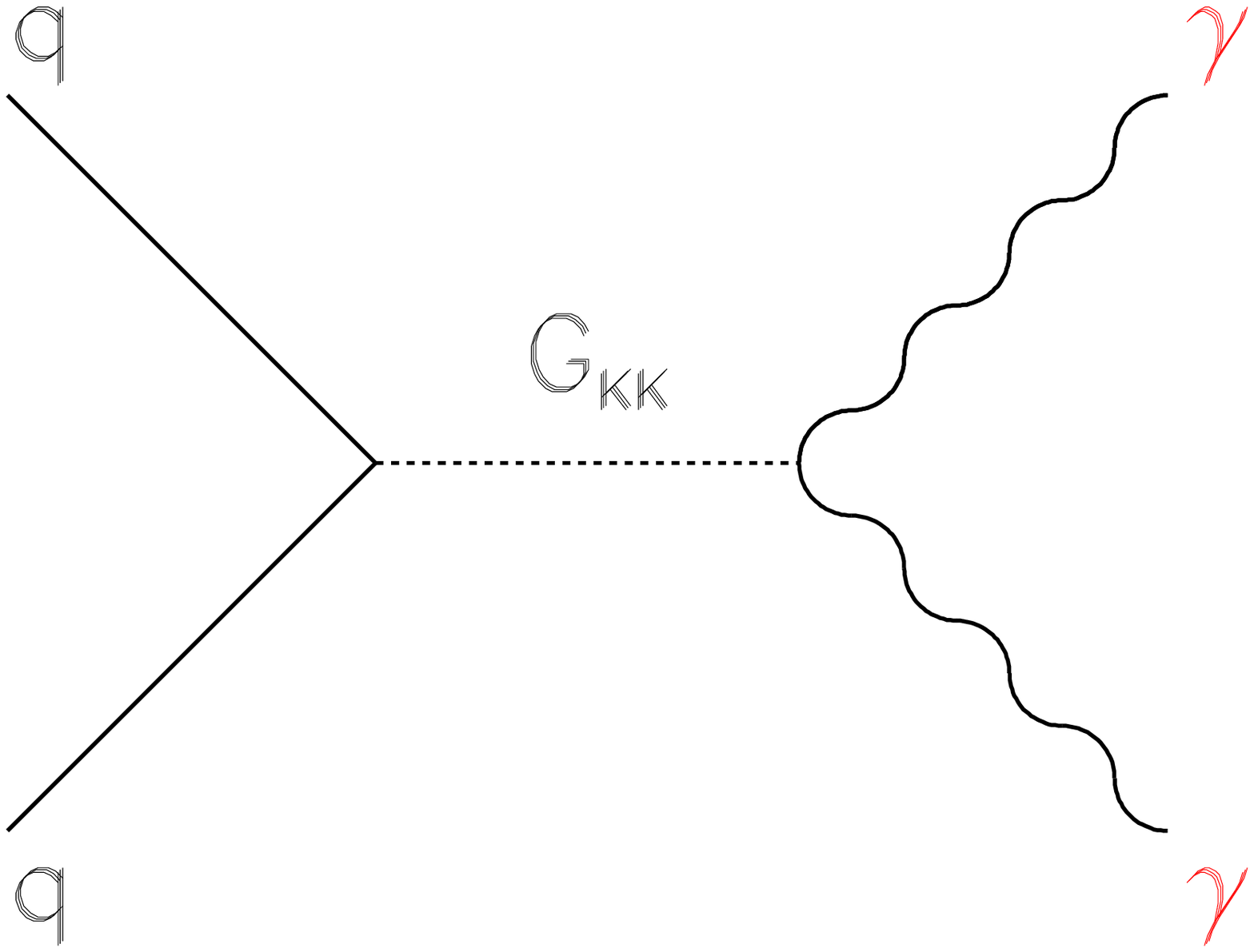, width=3.3cm}&
\epsfig{file=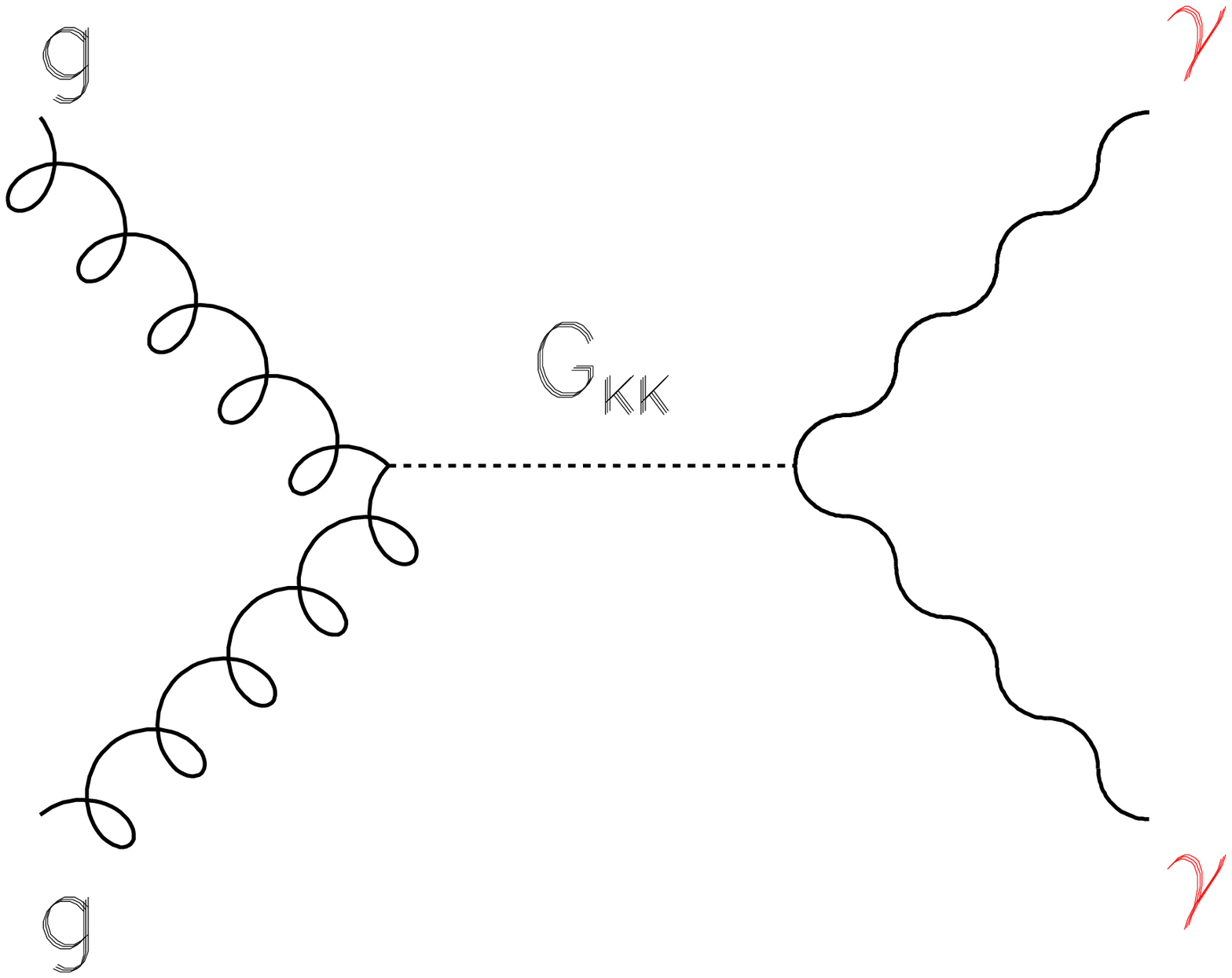, width=3.3cm} 
\end{tabular}
\parbox{15cm}{\caption{{\small {Diagrams of the standard model and KK graviton
contributions to di-photon production.}}} \label{diphoton}}
\end{center}
\end{figure}

Two types of scattering processes have been
investigated at the Tevatron Run\,1: the di-lepton production and the
di-photon production. All diagrams pertaining to these processes
are displayed in Figures~\ref{dilepton} and~\ref{diphoton}. These two
processes can be treated either inclusively (\Dzero\ analysis) or
exclusively (CDF analysis). In both cases, the novel processes interfere with
the standard model ones and the effective differential
production cross section requires an explicit cutoff, naturally at
\ms, due to the divergence of the sum over KK states,
and can be written:

\begin{equation}
\frac{{\rm d}^2\sigma}{{\rm d} \cos \theta^* {\rm d} M_{\ell\ell,\gamma\gamma}}=
\frac{{\rm d}^2}{{\rm d} \cos \theta^* {\rm d} M_{\ell\ell,\gamma\gamma}}
(\sigma_{SM}+\sigma_4 \eta+\sigma_8 \eta^2)
\end{equation}

where $\eta\equiv{\mathcal F}/M_S^4$ and $\sigma_{SM}$, $\sigma_4$,
$\sigma_8$ are respectively the standard model, interference and KK
individual contributions. M$_{\ell\ell,\gamma\gamma}$ and $\theta^*$
are respectively the invariant mass and the decay angle of the
two leptons or photons. The ${\mathcal F}$ factor embeds the
dependence on the choice of formalism. Three specific cases are
considered. In the first (Hewett scheme~\cite{hewett}) neither the
dependence on the sign of the interference ($\sim \lambda$) nor on the
number of extra dimensions are determined, ${\mathcal
F}=2\lambda/\pi$. In the second (so-called GRW~\cite{grw} formalism)
the sign of the interference is specified and ${\mathcal F}=1$. In the
third (HLZ formalism~\cite{hlz}) both dependences are accounted for,
and ${\mathcal F}=\log(M_S^2/s)$ for $n=2$ and ${\mathcal F}=2/(n-2)$
for $n>2$.

\vspace{.5cm}
\noindent{\it Inclusive di-electron and di-photon production at \Dzero}
\vspace{.5cm}

The study of a di-electromagnetic-particle (EM) production (electrons
or photons) at \Dzero\ was motivated by the gain in the discovery
potential from the combination of the two final states.  Muons were
not considered due to the poor high muon momentum resolution. A
specific leading order parton level generator was designed to simulate
the expected signal and its interference with the standard model
process~\cite{greg} (using {\it CTEQ4-LO}~\cite{cteq} parton
distribution functions). Initial state radiation is simulated
according to Drell-Yan (DY) measurements and the DY K-factor is
used. The main backgrounds to this analysis are the DY di-electron
production which is determined from Monte Carlo simulation and
instrumental backgrounds originating from di-jet production or
Compton-QCD processes where one or both jets are mis-identified as
electromagnetic particles. All other backgrounds such as W boson
production with a mis-identified jet, WW, $\mathrm{t}\bar\mathrm{t}$
into di-electron, Z into a pair of taus or Z$\gamma$ contribute to
less than 1\%\ to the total.

\begin{figure}[t] 
	\begin{center}
	\begin{tabular}{ccc}
	\epsfig{file=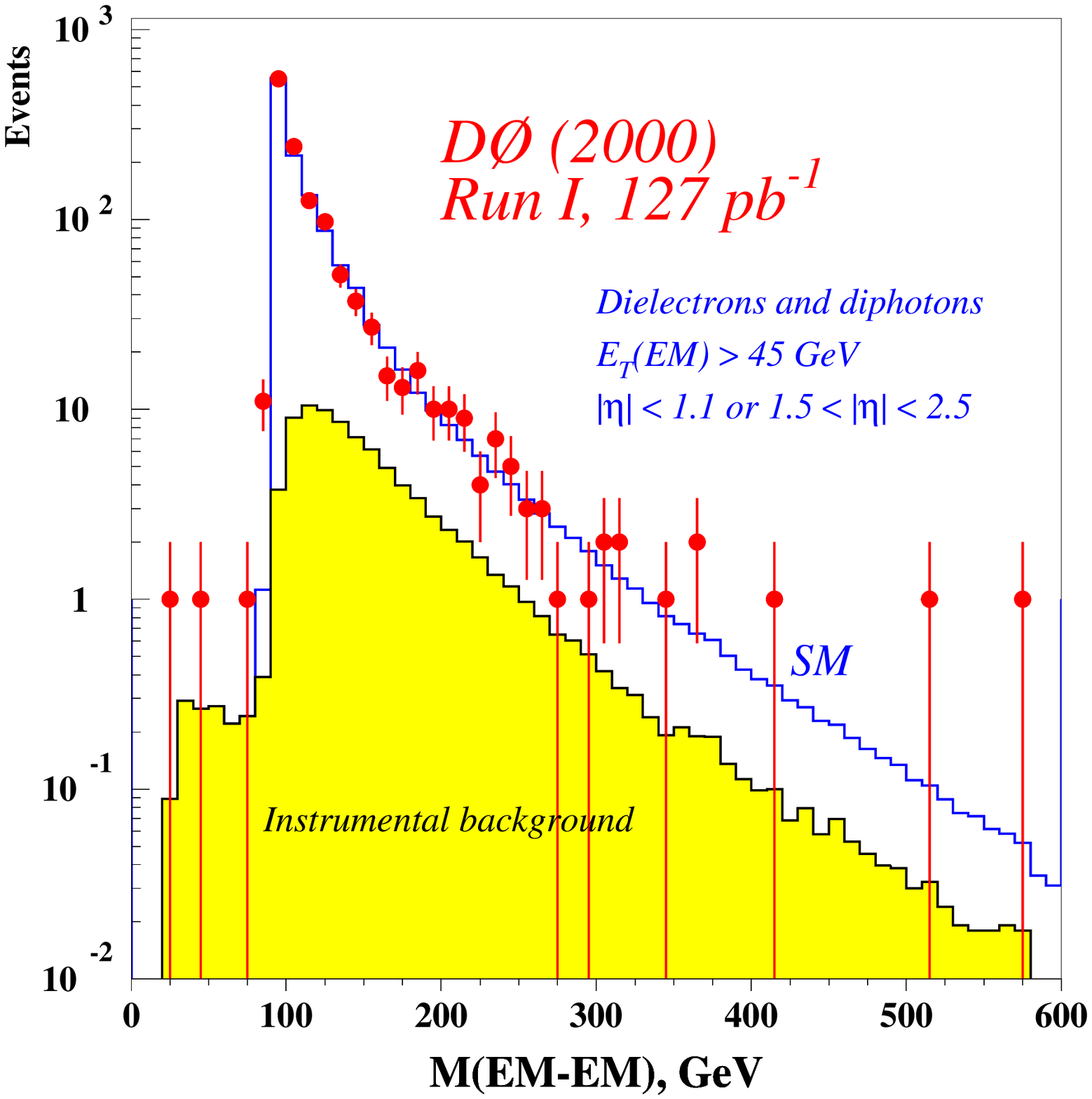, width=8cm, height=6cm} \hspace{-.5cm} &
	\epsfig{file=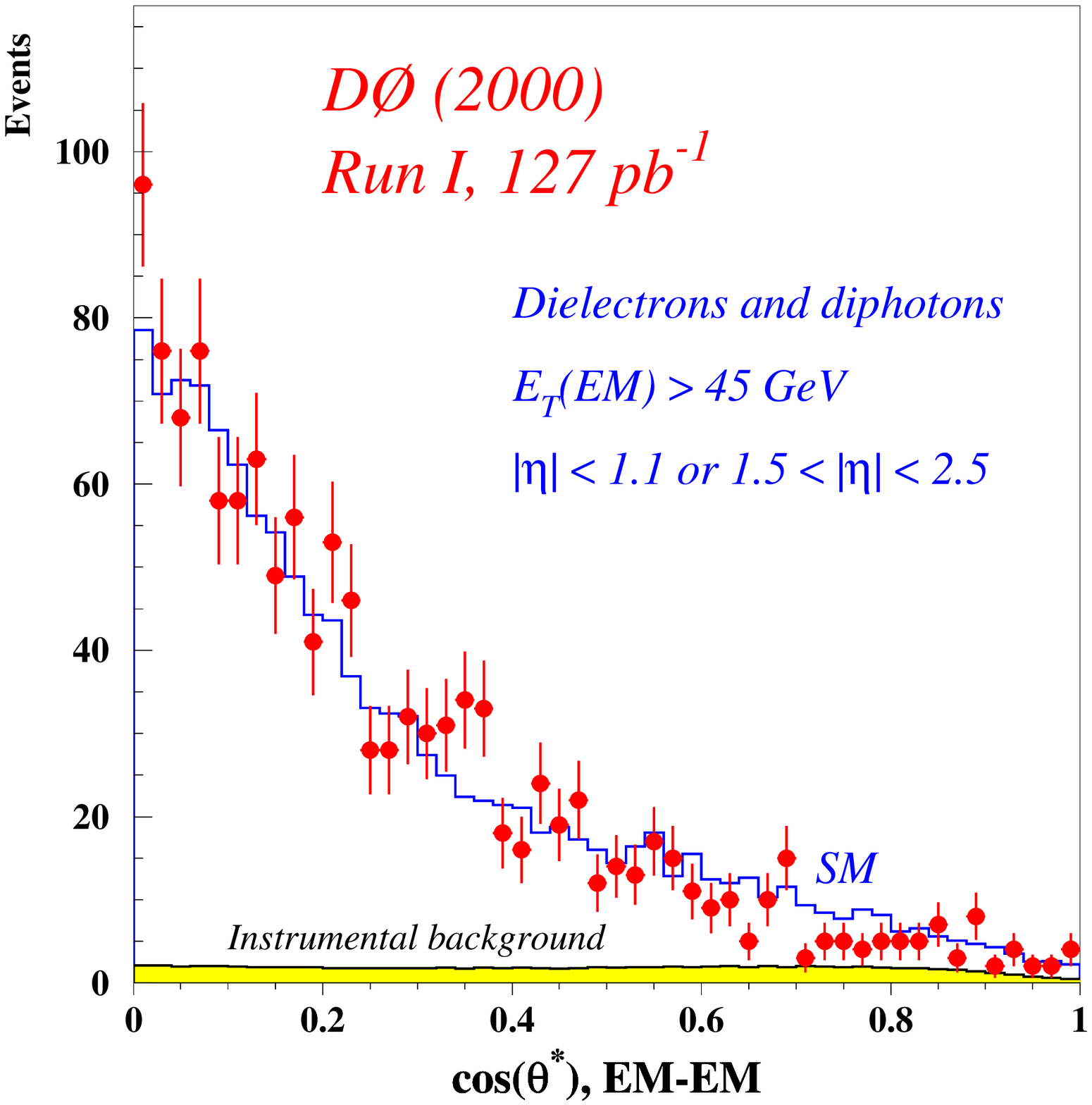, width=8cm, height=6cm} 
	\end{tabular}
  \parbox{15cm}{\caption{{\small {Distribution of the invariant mass
  (left) and the decay angle (right) of the two EM particles
  for data (dots with errors) and the background expectation
  (histogram). The instrumental background contribution is also
  indicated (shaded). }}} \label{diem-distributions}}
\end{center}
\end{figure}

\begin{figure}[t] 
	\begin{center} \epsfig{file=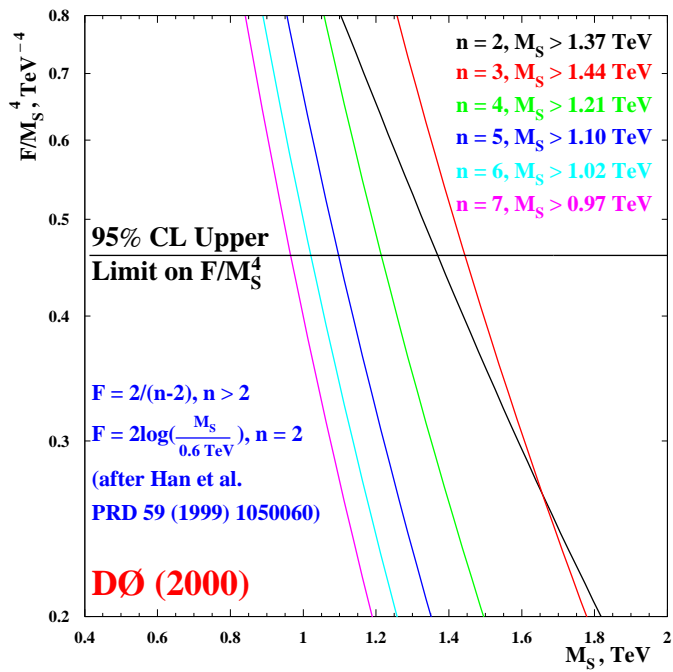, width=10cm}
	\parbox{15cm}{\caption{{\small {Limits on the Planck scale
	$M_S$ as a function of $\eta$ and the number of extra
	dimensions $n$ from the inclusive di-electron and di-photon
        search in \Dzero.}}}  \label{diem-limits}}
\end{center}
\end{figure}

The analysis is based both on the invariant mass and the decay angle
of the EM particles.  These two discriminant variables are shown in
Figure~\ref{diem-distributions} and are combined in a two dimensional
binned likelihood which takes into account systematic uncertainties
(These are dominated by the uncertainty originating from the
evaluation of next-to-leading order effects taken into account
via a K-factor). The result of these searches interpreted in the HRZ 
formalism is shown in Figure~\ref{diem-limits}. In the Hewett scheme, 
the limits
are $M_S > 1.1$\,TeV and $M_S>1.0$\,TeV respectively for $\lambda=1$ and
$\lambda=-1$. In the GRW formalism, $M_S>1.2$~TeV.

\vspace{.5cm}
\noindent{\it Exclusive di-electron and di-photon production at CDF}
\vspace{.5cm}

Because of its tracking capabilities within a magnetic field, in Run\,1 the
CDF experiment had a better discovery potential in the electron
channel and therefore does not perform an inclusive treatment of
electrons and photons. A preliminary analysis taking only into account
the invariant mass information was performed and yielded results
similar to those obtained by \Dzero. In the Hewett scheme, the limits
are \mbox{$M_S > 0.94$\,TeV and (resp.~0.85~TeV) for $\lambda=1$ (resp.~$\lambda=-1$)}.

\vspace{.5cm}
\noindent{\it Kaluza-Klein Graviton Emission}
\vspace{.5cm}

\begin{figure}[t] 
	\begin{center}
		\begin{tabular}{ccc}
 	 	\epsfig{file=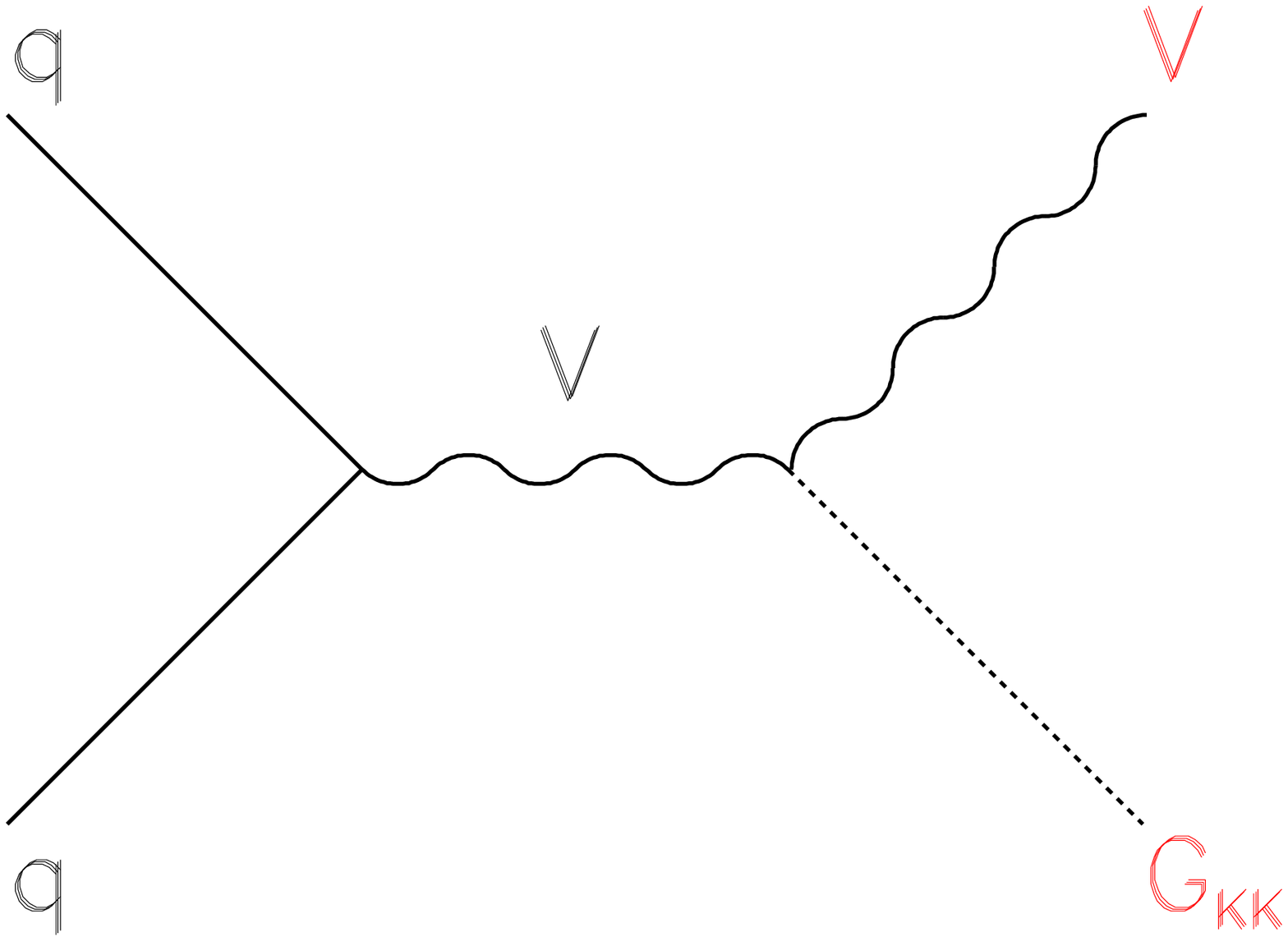, width=3.3cm}&
 	 	\epsfig{file=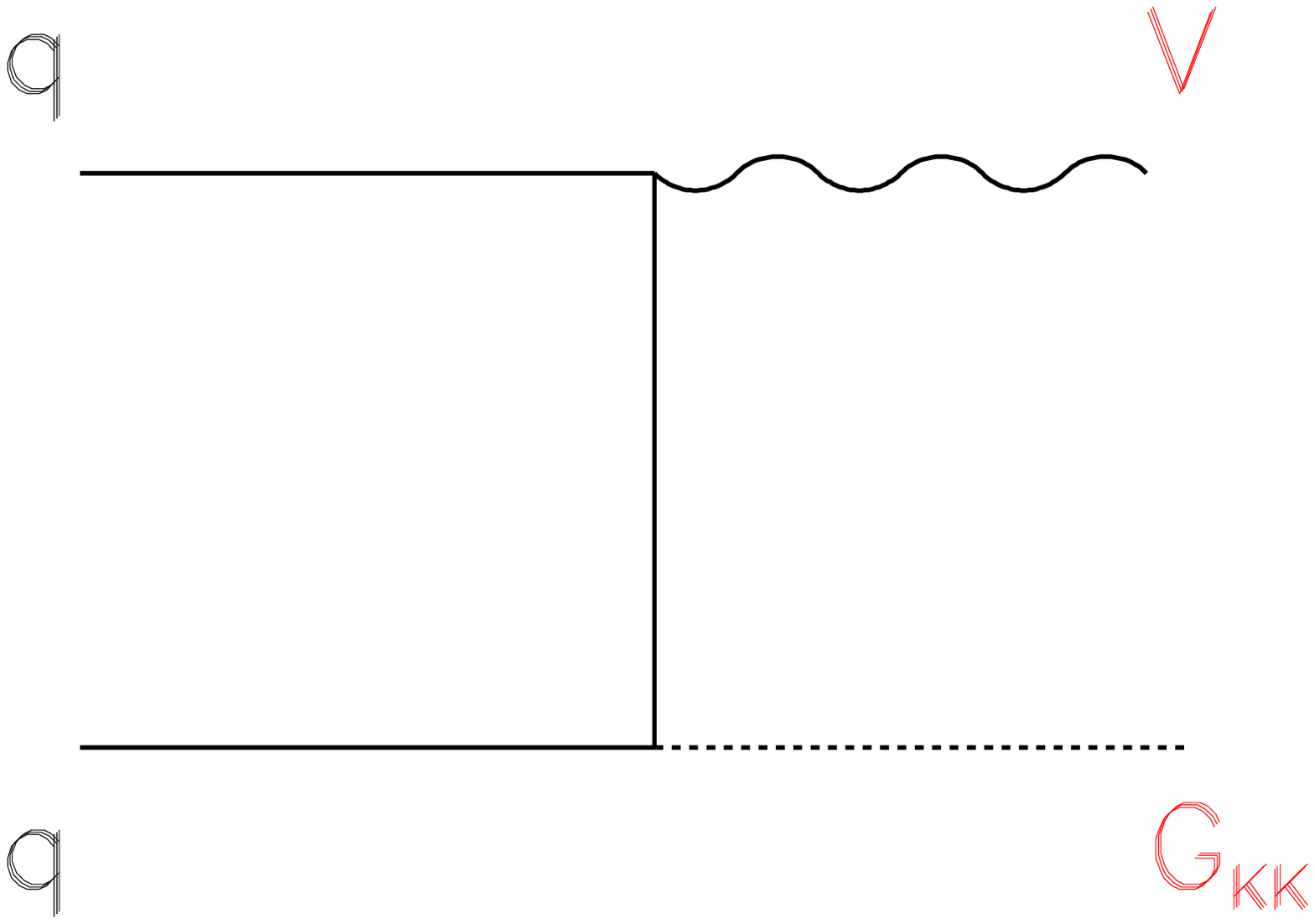, width=3.3cm}&
 	 	\epsfig{file=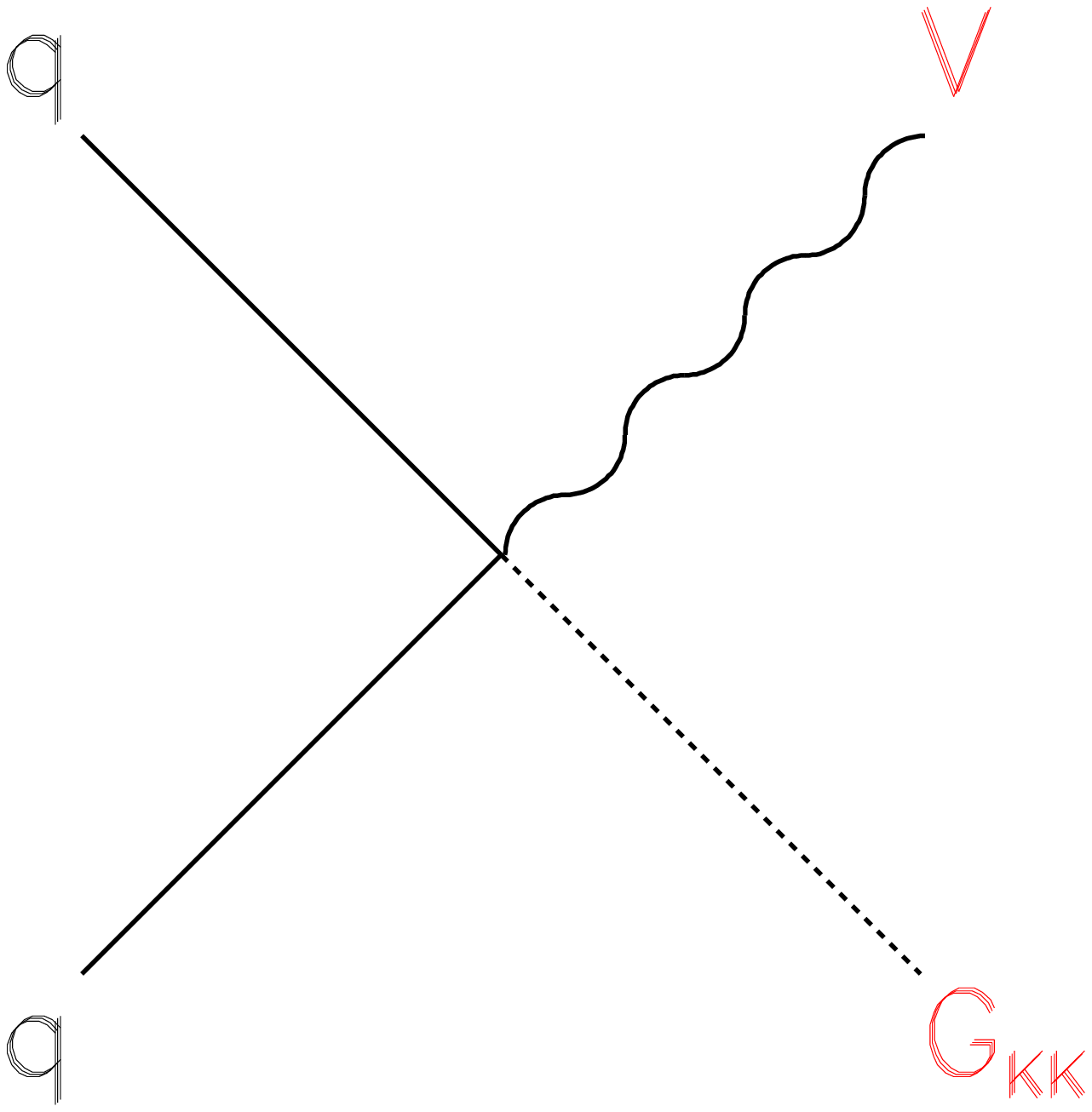, width=3.3cm}
		\end{tabular}
		\begin{tabular}{cc}
 	 	\epsfig{file=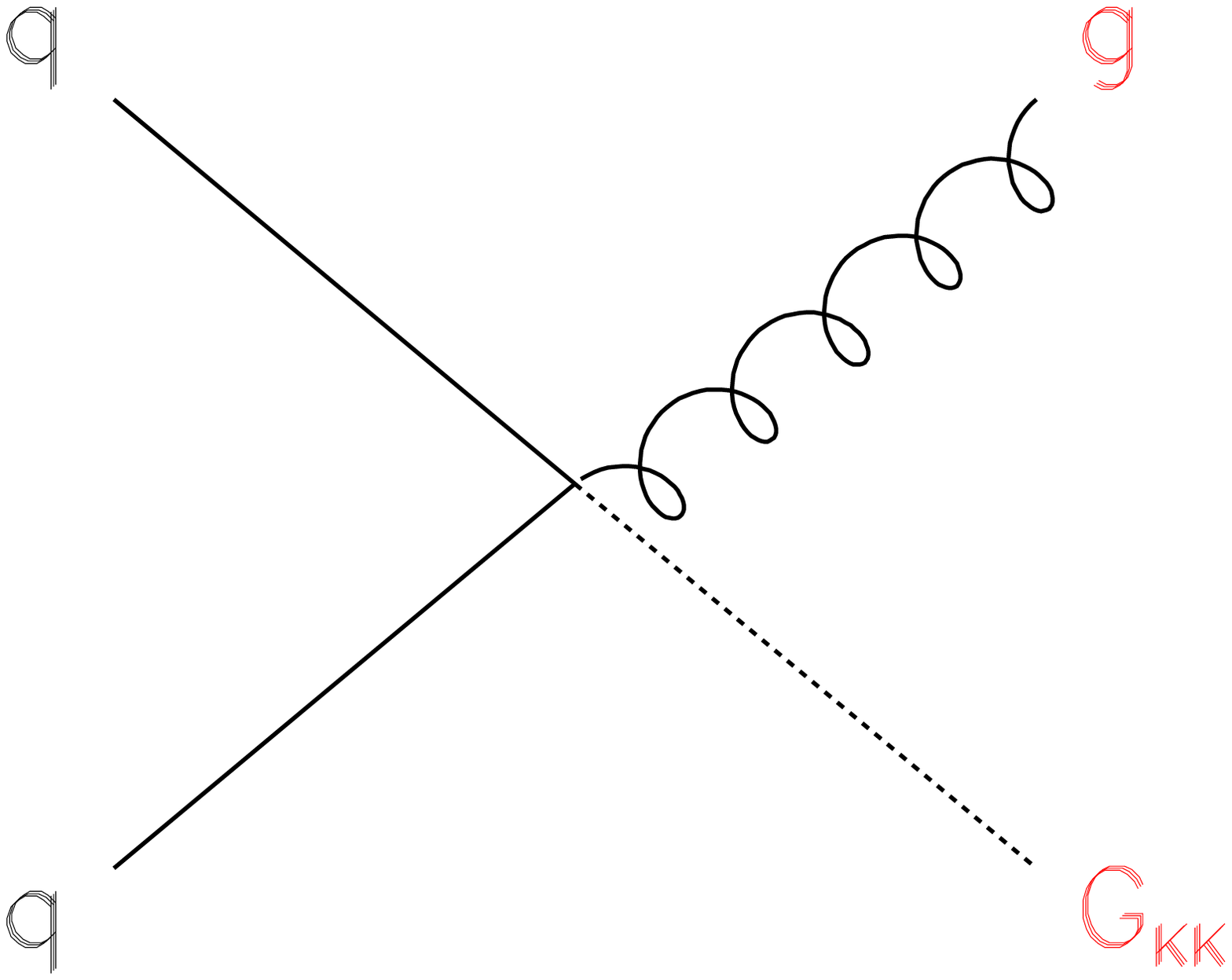, width=3.3cm}&
 	 	\epsfig{file=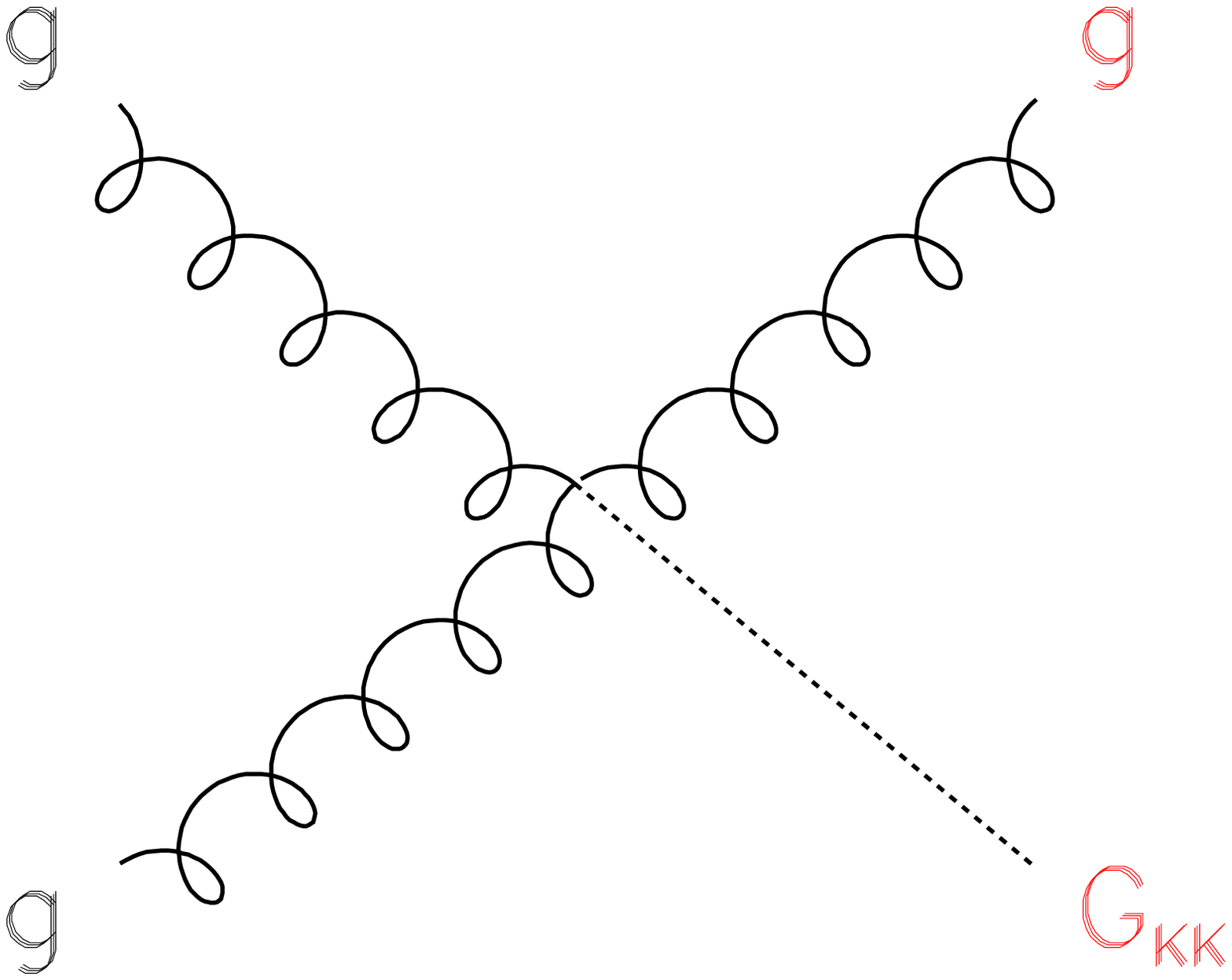, width=3.3cm}
		\end{tabular}
  \parbox{16cm}{\caption{{\small {Single vector boson production in association with a KK graviton.}}}\label{monojet-diagrams}}
\end{center}
\end{figure}

Direct evidences of KK gravitons can be searched for at the Tevatron
in their production in association with a vector boson. Since
gravitons escape detection, their distinctive signature is missing
transverse energy. The tree level direct KK graviton production
diagrams are depicted in Figure~\ref{monojet-diagrams}. Among all
possible final states arising from these processes, two have already
been investigated. CDF has performed an analysis of single photons and
missing transverse energy. \Dzero\ searched for mono-jet events.

\vspace{.5cm}
\noindent{\it The CDF single photon search}
\vspace{.5cm}

KK gravitons produced in association with a single photon were
searched for by the CDF experiment.  This search was performed on a
sample of $\sim 90$\,pb$^{-1}$. In this analysis only very clean events
where one photon with transverse energy in excess of 55\,GeV within a
pseudo-rapidity of $|\eta^\gamma|<1$ and a missing transverse energy
larger than 45\,GeV are required. Events where a jet with transverse
energy larger than 15\,GeV or a track with transverse momentum in
excess of 5\,GeV/c are removed. The largest background contribution
($\sim 60$\%) is due to cosmic rays where the muon undergoes a
bremsstrahlung in the Central Electromagnetic Calorimeter. Another
substantial background ($\sim 30$\%) process is the Z$\gamma$
production when the Z boson decays to a pair of neutrinos.

Eleven events were selected in the data in good agreement with the
expected background estimation of 11$\pm 2$ events. Limits on the
effective Planck mass are thus derived for various numbers of extra
dimensions (n):

\begin{equation}
\left\{ \begin{array}{c}
	{\hbox{n=4,     } \;\;  M_S>550 {\hbox{ GeV}}} \\
	{\hbox{n=6,     } \;\;  M_S>580 {\hbox{ GeV}}} \\
	{\hbox{n=8,     } \;\;  M_S>600 {\hbox{ GeV}}} 
	\end{array}
\right.
\end{equation}

\vspace{.5cm}
\noindent{\it The \Dzero\ mono jet search}
\vspace{.5cm}

The topology searched for in the case of a KK graviton produced in
association with a gluon is a single jet. A search for such monojet
events was carried out by the \Dzero\ experiment with a data sample of
$\sim 80$\,pb$^{-1}$. The event selection requires one or two jets,
where the leading jet must have a transverse energy in excess of
150\,GeV and the second jet should not have a transverse energy larger
than 50\,GeV and should not be pointing in the azimuthal direction of
the missing transverse energy in order to suppress the QCD
background. The most prolific background contribution is that of W and
Z production which is expected to yield 30$\pm$4 events. From QCD and
cosmic backgrounds 8$\pm$7 events are expected to be produced.

\begin{figure}[t] 
	\begin{center}
	\epsfig{file=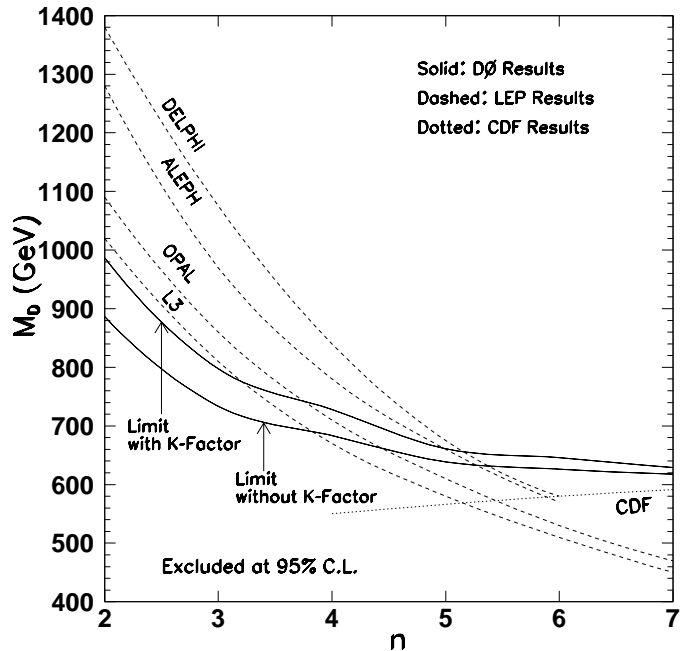, width=10cm}
  \parbox{15cm}{\caption{{\small {Limits on the Planck scale $M_S \sim M_D$ as a function of the number of extra dimensions from the monojet search in \Dzero.}}}  \label{monojet-limits}}
\end{center}
\end{figure}

Altogether, 38$\pm$8 events are expected from standard model processes
and cosmic rays, in good agreement with the 38 events observed. Limits
on the Planck scale can thus be derived as a function of the number of
extra dimensions, as illustrated in Figure~\ref{monojet-limits}. These
limits include the next-to-leading-order correction embedded in
the K factor. For n=4 and n=6 the limits are respectively
$M_S>698$\,GeV and $M_S>632$\,GeV. As shown in
Figure~\ref{monojet-limits} the limits obtained by this analysis
improve the exclusion with respect to the LEP limits only at large
number of extra dimensions.

\vspace{.5cm}
\noindent{\bf {\large Warped extra dimensions (RS)}}
\vspace{.5cm}

\begin{figure}[t] 
	\begin{center}
	\mbox{ 
	\hskip -1.0cm
	\epsfig{file=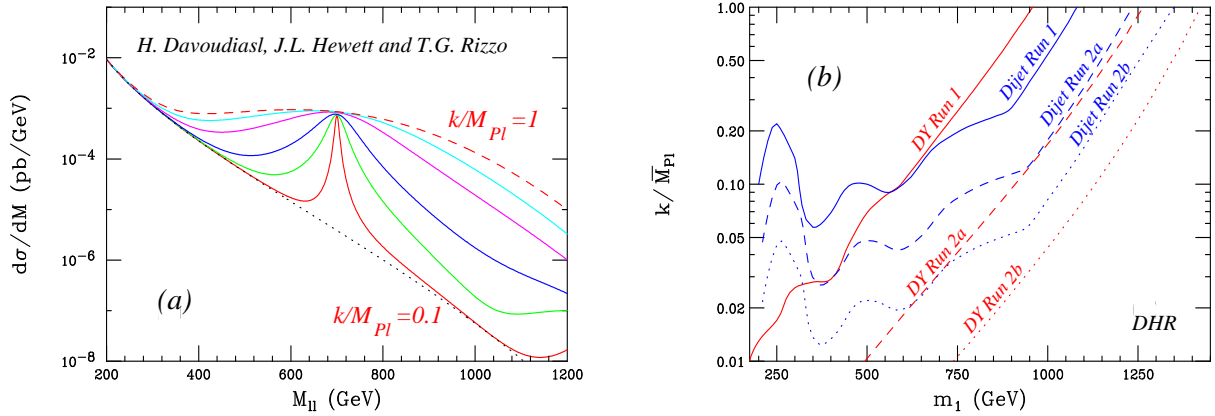, width=18cm}}
	\parbox{15cm}{\caption{{\small {(a) Production of a 700~GeV/c$^2$
	first KK graviton for various values of k/$M_{Pl}$=1, 0.7,
	0.5, 0.3, 0.2 and 0.1 from top to bottom respectively. (b) Limits on $k/M_{Pl}$ as a function of $m_1$.}}}	
\label{rsprod}}

\end{center}
\end{figure}

In the Randall-Sundrum scenario, the effective hierarchy is explained
by an exponential warp factor in a 5-dimensional Anti-de-Sitter
(AdS$_5$) geometry. The single extra dimension in this scenario is
finite and of size $y=\pi R_c$ (where $R_c$ denotes the
compactification radius). The metric preserving the 4-dimensional
Poincar\'e invariance:
\begin{equation}
ds^2=e^{-2ky}\eta_{\mu\nu}dx^\mu dx^\nu - dy^2
\label{rsmetric}
\end{equation}
is considered. The exponential is known as the warp factor and the
parameter $k$ is the AdS$_5$ curvature scale, expected to be of the
order of M$_{Pl}$. Since the scale of the physical processes appears
to be $M_{Pl}e^{-ky}$, the hierarchy between the TeV and the Planck
scale arises naturally when gravity is localized on the $y=0$ brane
and all physical processes occur on the $y=\pi R_c$ brane with $k R_c
\simeq 11-12$ (TeV-brane). Here it is assumed that the standard model
fields are confined to the TeV brane. The sole two excitations arising
from the metric of eq.~\ref{rsmetric}, {\it i.e.} the usual graviton
related to the 4-d Minkowski metric and the so-called radion related
to the transverse 5$^{th}$ dimension (the relative motion of the two
branes), are the only ones allowed to propagate freely in the bulk.
The phenomenology of this model is very broad~\cite{rs2}. The
particular case of the radion has mostly been investigated in the
framework of the LHC~\cite{lhctalk}, it is therefore not discussed
here. However, in this scenario the graviton production yields a very
distinctive and attractive signature at hadron colliders, since the KK
modes are not evenly spaced and are separated well enough so that a
resonant production can be observed. The first KK state is
naturally of the order of a TeV. The production cross section of a
700\,GeV first KK state at the Tevatron is shown in
Figure~\ref{rsprod}a for various values of $k/M_{Pl}$.

Using the Tevatron Run\,1 analyses of the Drell-Yan spectrum in search
for heavy gauge bosons and of the Dijet production, a limit on
$k/M_{Pl}$ as a function of the mass of the first KK state ($m_1$) is
derived~\cite{rs2}. These results are displayed in
Figure~\ref{rsprod}b.

\vspace{.5cm}
\noindent{\it Prospects and status of Run\,2}
\vspace{.5cm}

The Run\,2a of the Tevatron has successfully started in spring
2001. As of July 2002, the highest peak luminosity reached was
$2\,10^{31}$~cm$^{-2}$s$^{-1}$ and the integrated luminosity delivered
was of the order of 50\,pb$^{-1}$. Although at the time of the
conference the detectors were still in a commissioning phase, first
physics results started to be produced. With the increase in
center-of-mass energy and luminosity, the sensitivity to the gravity
scale in ADD models is expected to double at Run\,2a with respect to
Run\,1 and triple at Run\,2b.  Other final states, such as those
containing muons, can be searched for with the improved detector
capabilities. Run\,2 prospects of the searches for extra dimensions in
the RS scenario are displayed in Figure~\ref{rsprod}b. These results
are obtained by rescaling the Run\,1 result accounting for the
increase in luminosity and the expected variation of the cross section
due to the increase in center-of-mass energy~\cite{rs2}.

\vspace{.5cm}
\noindent{\bf {\large Conclusions}}
\vspace{.5cm}

The CDF and \Dzero\ experiments at Tevatron Run\,1 have searched for
signs of extra spatial dimensions. No excess has been observed in any
of the channels studied. The observed exclusion limits are of the same
order as those obtained at LEP~\cite{leptalk}. The Tevatron Run\,2 has
successfully started its data taking phase and will provide an
enticing potential for discoveries in this exciting field.

\newpage

\vspace{.5cm}
\noindent{\bf {\large Acknowledgements}}
\vspace{.5cm}

We wish to thank Robert Zitoun for his careful reading of these proceedings.

\vspace{.5cm}
\noindent{\bf {\large References}}
\vspace{-1.3cm}


\end{document}